# Graphene Oxide flakes: methods and techniques for properties at interfaces

**Zuppella P.[1], Zuccon S.[1], Nardello M.[1, 3], Corso A.J.[1], Silvestrini S.[2], Maggini M.[2] and Pelizzo M.G.[1, 3]**

1. CNR–IFN UOS Padova, Via Trasea 7, Padova
2. University of Padova, Department of Chemical Sciences, Padova, Italy
3. University of Padova, Department of Information Engineering, Padova, Italy

**Abstract.** Graphene Oxide and reduced Graphene Oxide are intriguing materials for photonics and electronic devices both for intrinsic characteristics and as precursors for the synthesis of graphene. Whatever the application and the engineering purpose, a fine control of the chemical and physical properties is required since the performances of graphene–based systems depend on the reduction state of Graphene Oxide and can be strongly affected by interfaces interactions and neighboring effects. Then, a method for a local control of electric, electronic and chemical properties is required. The synergic application of Near–Edge X ray Absorption Fine Structure Spectroscopy and Conductive–Atomic Force Microscopy is a powerful way for such purpose allowing a full characterization of simple and composite samples, including specimens with dielectric discontinuities. Graphene Oxide/Gold (GO/Au) and Graphene Oxide/Silicon Oxide (GO/SiO$_2$) have been selected as proof by example. The results show that the approach allow to relate chemical, electronic and physical properties to morphological features and local conductance behavior.

**1. Introduction**

First synthesized almost 150 years ago, GO has been mainly considered as a precursor for cost–effective, large–scale production of graphene–based materials [1, 2]. It has recently attracted attention for many applications due to its insulating, transparent, and hydrophilic properties. Layers of GO have a large amount of oxygen functional groups decorating the basal plane and the edges of a typical graphene sheet. Most of the carbon atoms bonded with oxygen is $sp^3$ hybridized and disrupts the extended $sp^2$ conjugated matrix of original graphene with a consequent higher interlayer distance compared to that of graphite. The conspicuous $sp^3$ fraction in GO is responsible of its insulating behaviour even though incremental removal of oxygen can move the material to a semiconductor and ultimately to a graphene–like semimetal [3]. Substrate dependent effects can affects the opto–electronic properties, then the knowledge of GO/thin films interfaces represents a nontrivial challenge for many applications and for optimal engineering purposes. Interfacial hybridization and bonding can potentially enhance or decrement electronic properties by inducing charge redistribution, introducing hybrid states within the band structure depending upon the nature of chemical bonding at the interface. A lot of work both experimentally as well as theoretically has been done even in a short span of time to better understand the fundamental and applied science of graphene–based material, but many aspects still remain to investigate and a bottom–up approach is often required [4, 5]. Whatever the application, a proper methodological approach must be identified. For example, a local conductance analysis and its relation to the chemical bonding at the interfaces is needed at nanoscale for a detailed study of the electronic transport phenomena [6–8]. Conductive–Atomic Force Microscopy (C–AFM) has been recently applied for nanoscale conductance scanning of graphene layer [9, 10]. The C–AFM is well suited for a simultaneous mapping of morphology and local electrical discontinuities with a nanoscale resolution. One of the advantages is that it can be used unlike a traditional Scanning Tunneling Microscopy (STM), for measuring specimens showing dielectric regions. The technique can be combined with the Near–Edge X ray Absorption Fine Structure Spectroscopy (NEXAFS), well established to characterize element–specific electronic structure with extremely high surface sensitivity. NEXAFS is useful spectroscopy for probing surface chemistry, surface molecular orientation, degree of order, and electronic structure of nanomaterials [11–15]. The peak positions and lineshapes of the observed NEXAFS resonance represent to first approximation, a replica of the unoccupied atom–projected density of states modified by core–hole interactions. GO/Au and GO/SiO$_2$ chips have been selected as test samples together with a commercial Graphene/Ni assumed as reference. NEXAFS spectroscopy together with Scanning Probe Microscopies and local conductance mapping has been successfully applied to the specimens in order to fully characterize the GO flakes on the surfaces. The results and the analysis allow to relate morphological features, chemical properties and electrical behavior at nanoscale, confirming that the synergic use of both techniques allows a deep knowledge of such samples.

**2. Experimental**

The pristine samples consist of 20 nm Au and 100 nm $SiO_2$ deposited respectively onto two Silicon wafers by Electron Beam. A protocol based on the sonication in Isopropyl alcohol (IPA) and heating of the specimens has been applied for cleaning and drying them. Several GO sheets dispersion are commercially available. We used Graphene Supermarket Dispersion in water of Single Layer Graphene Oxide, 500 mg/L concentration and 0,3 – 0,7 μm flakes size. The solution has been further diluted in IPA (1 part of GO and 4 parts of IPA) and then deposited onto the samples by spin coating method including three steps at 300 rpm/s, 400 rpm/s and 500 rpm/s for a total of four minutes. First observations by optical microscope allowed to verify macroscopic features and defects on the surface.

A multi–location morphological scanning of the samples has been performed in ambient atmosphere by using XE–70 AFM Park System in Contact, Non–Contact (NC–AFM) and Conductive (C–AFM) modes, depending on the samples. In NC–AFM, the tip is held immediately above the surface; it measures surface topography via deflections caused by longer–range attractive interactions. The absence of repulsive forces allows to scan "soft" samples reducing the risk to break them. The C–AFM simultaneously records topography and current distribution over the surface. A bias voltage ranging from 10 V to -10 V can be applied between sample and tip (ground reference) and the tunneling current between them is measured. While the topography is acquired using the deflection signal of the cantilever, the electric conductivity is measured through an electric current amplifier. The electric current through the cantilever can be as small as pA, then the current amplifier is chosen so that the electrical noise can be suppressed up to fA. We selected a conductive tip PtIr coated, $10^9$ V/A gain giving few pAs noise values, appropriate for the conductive mapping of GO – Au sample. The images in the following section are 2,5 μm x 2,5 μm scan areas; the mappings have been performed in the regions investigated by NEXAFS spectroscopy.

NEXAFS analysis are well established for obtaining a refined map of the unoccupied density of states (UDOS), thereby providing a means to describe element–specific electronic structure with extremely high surface sensitivity. The close correlation between NEXAFS data and UDOS depends on the X–ray absorption cross–section in turn related to the energy density of the final states. The NEXAFS spectra shown in this paper have been recorded at Carbon–K edge; this kind of measurements needs Synchrotron Light to be performed. BEAR beamline–ELETTRA Synchrotron light source in Trieste [16] is equipped with a planar laminar grating with 1200 lines/mm, that allows to acquire the C K–edge spectra yielding an energy resolution of approximately 0,1 eV. The spot size on the sample is usually around 30 μm x 100 μm. The selected data have been acquired in Total Electron Yield (TEY) mode with a nearly linear polarization vector parallel to the plane of the samples. Many areas of the surfaces have been investigated. We report the most representative results together with the spectrum of a commercial Graphene/Ni chip showing the peculiar peaks of graphene and assumed as reference.

## 3. Results and discussions

The NEXAFS spectral features of graphene based materials in the regions 283–289 eV and 289–315 eV are ascribed to C $1s \rightarrow \pi^*$ and C $1s \rightarrow \sigma^*$ transitions, respectively [11, 12]. Fig. 1 shows the analysis of Graphene/Ni commercial specimen. The top layer of the sample consists of tens of nanometers of graphene as it is shown in AFM measurements (Fig. 5, right), therefore the spectrum retraces the typical resonance peaks of graphite. The $\pi^*$ and $\sigma^*$ resonances are well marked and the pre–edge and putative interlayer/functionality features also appear at ~ 284 eV and ~ 290 eV [12].

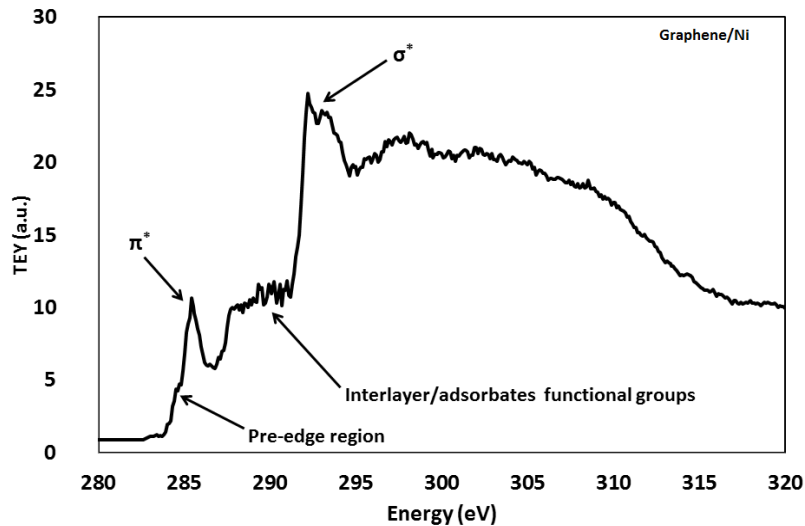

Fig. 1 C K–edge spectrum of commercial Graphene/Ni. The NEXAFS spectrum reveals the typical $\pi^*$ and $\sigma^*$ resonances.

Going from graphene to GO, the C–K edge spectra are characterized by broadening and reshaping of the $\pi^*$ and $\sigma^*$ features as it has been found for suspended GO sheets revealing the GO fingerprints [12]. All the measurements, both for commercial and homemade samples, have been performed with a nearly linear polarization vector parallel to the plane (inset Fig. 2). The intensity and the shape of the peaks are determined by the dipole selection rules: an orbital with perpendicularly symmetric orientation to the basal plane ($\pi^*$) should not be resonant if the electric field vector of the incident beam is coincident with the nodal plane of the sheet. In the case of perfect $sp^2$ hybridization and ordered structure, considering the mutual orientation of the electric field and the samples (inset Fig. 2), the $\sigma^*$ states should be well defined and the $\pi^*$ peak should be suppressed, that doesn't happen even for the commercial sample. Instead, the sharp peak due to $\sigma^*$ states disappears in the case of GO/Au and softens for GO/SiO$_2$ while maintaining a clear ramp at 291,6 nm. Changes on the shape of $\sigma^*$ reveal the presence of defects and lower order in the structure. Peaks relaxed and not defined correspond to disorderly arrangements. There is a further difference between the GO/Au and GO/SiO$_2$ spectra around 290 eV. The GO/SiO$_2$ shows peaks that are usually attributed to interlayer interactions or adsorbates functional group. The existence of the $\pi^*$ features for both samples reveals corrugations and crinkling as well as $sp^3$ hybridized orbitals due to oxidation and defects [12]. The strong $sp^3$ hybridization is responsible of the insulating behaviour. The electrical properties of the flakes have been also confirmed by C–AFM measurements of GO/Au sample.

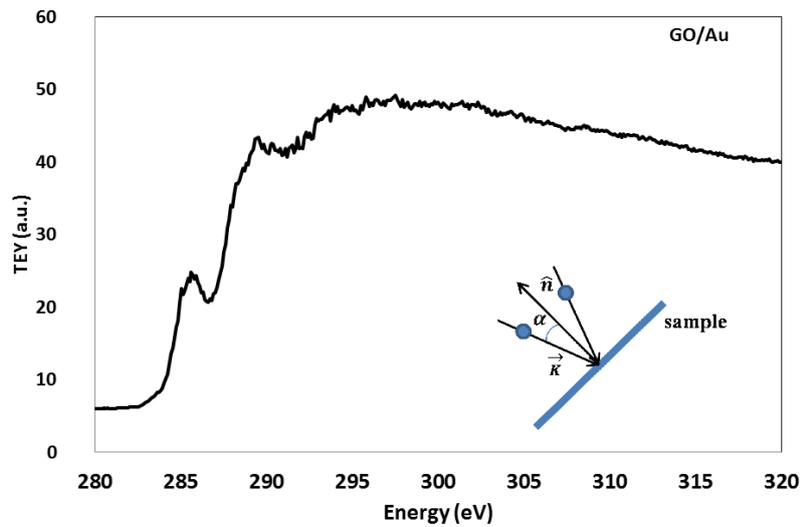

Fig. 2 C K–edge spectrum of GO/Au. All the NEXAFS spectra have been recorded by using a linear polarization beam probe as it is shown on the inset. The linear polarization vector is parallel to the plane of the samples.

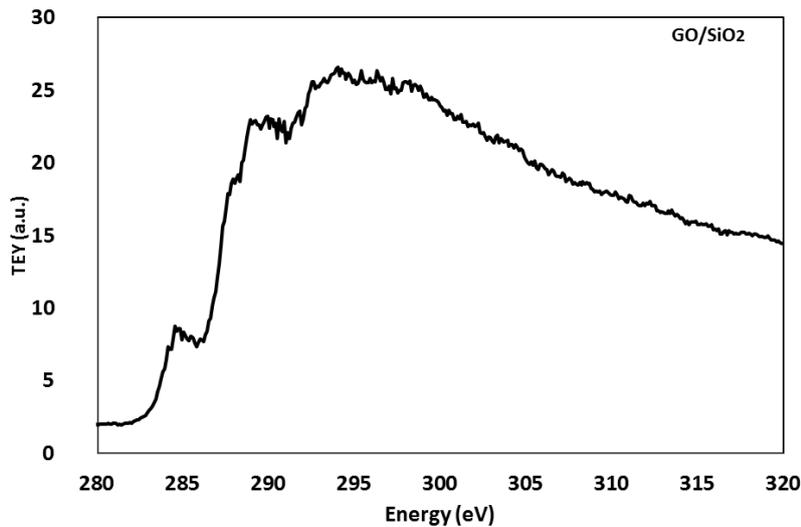

Fig. 3 C K–edge spectrum of GO/SiO$_2$. The broadening of the $\pi^*$ and $\sigma^*$ features reveals defects and lower order in the structure

The C–AFM is a powerful scanning microscopy able to relate morphological features and local conductance performances of the samples and it is well suited to investigate the local electronic properties of rGO based systems. The images obtained by using the setup previously described are topographic (Fig. 4, left) and current mapping (Fig. 4, right) of GO/Au chip. As observed in Fig. 4, the topography shows many domain structures: variable sizes GO flakes induce metal–insulating discontinuity on the surface. The Scanning Tunneling Microscopy (STM) is not suitable for this kind of systems and the C–AFM is required. In fact, the use of force feedback decouples the regulation of the tip sample distance from the current mapping, allowing to investigate samples with not conductive regions. The image in Fig. 4 (right) shows the local conductance carried out by applying -8 V voltage in ambient atmosphere. Negative bias voltage are interesting because may induce electrochemical reduction locally changing the electrical transport properties of GO [10]. The net value of the voltage has been selected to test the limit of the technique for our samples enhancing the conductance contrast between metal and insulating area without inducing reduction process. We have also verified that repeated scans at the same voltage (-8 V) strongly modify the sample morphology. Measurements at higher values seriously damage the samples and the flakes on it. Flakes based systems can be really breakable and flimsy, then a nearly contact operation can damage the topography and affect the local electric contact. The line profile (Fig. 4) shows features of less than ten of nanometers height with corrugation of about 2 nm and the corresponding current profile identifies the conductive and insulating regions. The current signal due to gold layer is around -120 pA, while the GO flakes are not conductive corresponding to a zero current on the map. The horizontal dark line in the current image are attributed to unstable electric contact between the tip and the sample [9] and to possible starting detachments of small piece of flakes. Clearly, the absolute value of current in C–AFM depends on the cantilever and tip condition, which is very sensitive on the tip sharpness and contamination of the tip apex. The insulating nature of GO flakes already proved by NEXAFS spectra is here confirmed.

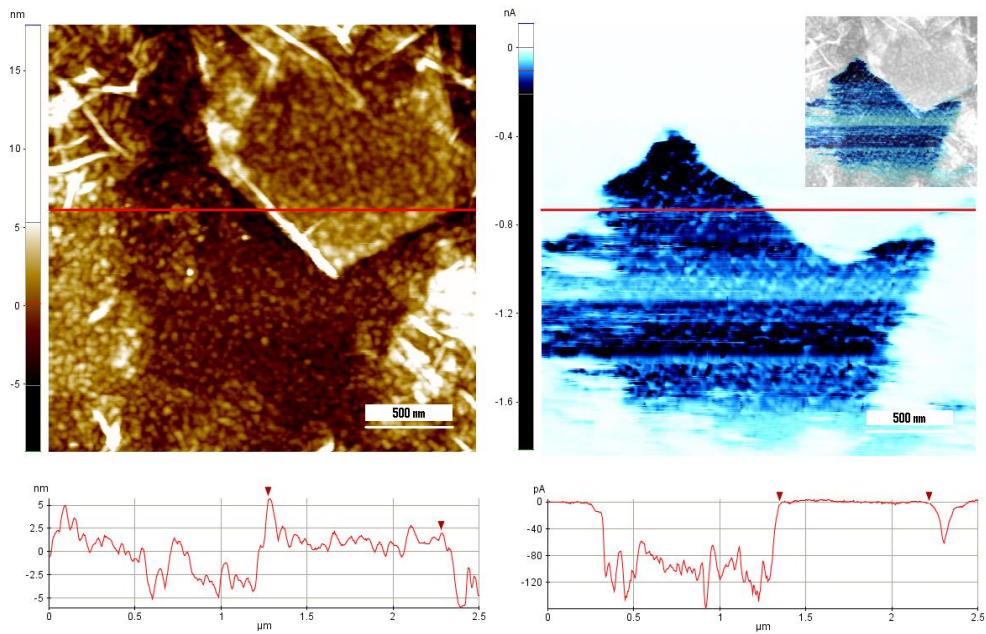

Fig. 4 Current mapping AFM results for few–layers GO onto Gold thin films. The measurement has been performed in contact mode with -8 V bias applied between sample and tip (ground reference). The current image on the right corresponds very well to the topography shown on the left clearly identifying the GO flakes. The line profiles were taken along the lines shown in images. The inset shows the overlap of current mapping and morphology revealing a good correspondence and the insulating essence of the GO flakes.

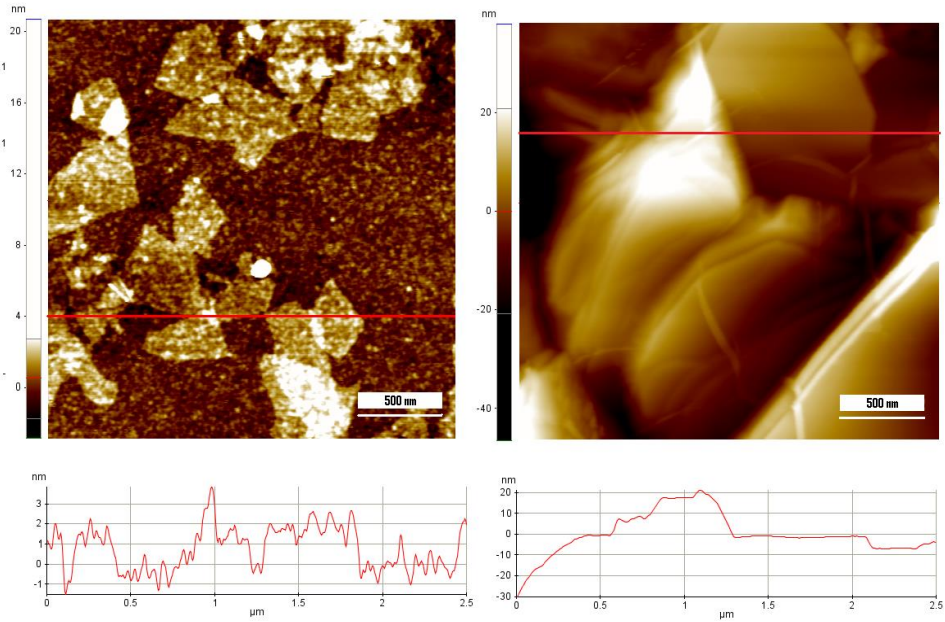

Fig. 5 (left) NC – AFM image of GO/SiO$_2$ sample. The morphology is patchy. The GO flakes are of undreds nanometers size and fews nanometers height; (right) Contact – AFM topography of graphene/Ni sample. The thickness of graphene is around tens of nanometers.

Fig. 5 shows the morphologies of the GO/SiO$_2$ sample and graphene/Ni commercial chip. The images have been recorded in NC–AFM and Contact mode respectively. From the morphological point of view, the grain size of the flakes on SiO$_2$ is less than micrometer with corrugations of the order of few nanometers, while tens of nanometers graphene cover the Ni on the commercial sample. In the present setup, the C–AFM cannot be applied to oxide layers unless to deposit a metallic electrode on the top. The conductive mapping of rGO/SiO$_2$ and GO/SiO$_2$ systems has been deeply studied [10], while the GO/Au system is still under investigation for a wide range of application [1].

**4. Conclusions**
The engineering of the surface properties and the nanoscale conductance of GO and rGO onto thin films is intriguing topic for photonic and electronic devices and flexible components. Therefore, methods to clarify the relationship between structure and electronic properties are required to be investigated. Au and SiO$_2$ have been selected as pristine materials to overlay with flakes of GO by spin coating technique. The test samples have been characterized by NEXAFS spectroscopy, Scanning Probe Microscopies and Conductive–AFM . The novel coupling of the techniques allow to relate chemical–electronic properties and electrical behaviour at nanoscale as it has been demonstrated in this paper.